\documentclass[aps,prc,preprint,groupedaddress]{revtex4-1}
\usepackage{color}
\usepackage{graphicx}

\begin{document}


\title{Investigating Local Parity Violation in Heavy-Ion Collisions Using Lambda Helicity}


\author{L.E. Finch}
\author{S.J. Murray}
\affiliation{Southern Connecticut State University, Department of Physics, New Haven, CT 06515}


\date{\today}

\begin{abstract}
We propose the measurement of net $\Lambda$ and $\bar{\Lambda}$ helicity, correlated event-by-event with the magnitude and sign of charge separation along the event's magnetic field direction, as a probe to investigate the Chiral Magnetic Effect in Heavy-Ion Collisions.  With a simple simulation model of heavy-ion events that includes effects of Local Parity Violation, we estimate the experimental correlation signal that could be expected at RHIC given the results of previous measurements that are sensitive to the CME.
\end{abstract}


\maketitle


\section{Motivation}

Measurements in heavy-ion collisions of charged particle azimuthal correlations have provided evidence for the effect known as Local Parity Violation (LPV) in which net topological charge in the collision environment generates a net chirality in an event \cite{Kharzeev:1998kz,Morley:1983wr}.  One manifestation of LPV is the Chiral Magnetic Effect (CME) \cite{Kharzeev:2007jp}, in which these topological effects combine with the large magnetic field generated by the colliding positive ions to produce a separation of electric charge along the direction of the event magnetic field.

Because the sign of the net topological charge is random event-to-event, the CME charge separation is not visible via any observable when averaged over many events and thus must be observed through particle correlations.  A typical measurement involves the correlation of particle directions taken with respect to the "reaction plane" (the plane containing the center of both colliding ions and the impact parameter) because this plane is highly correlated with the magnetic field direction in a collision.  In particular, the observable 
$\gamma=<cos(\phi_{\alpha}+\phi_{\beta})>$
was proposed as being sensitive to the Chiral Magnetic Effect \cite{Voloshin:2004vk}.  Here $\phi$ represents the azimuthal angle of a produced particle measured in a coordinate system in which the beam-direction is the z-axis and the x-z plane is the reaction plane, $\alpha$ and $\beta$ denote particle charges, and the average is taken over particles in an event, and then over events.  Both opposite-sign correlations ($\gamma_{+-}$) and same-sign correlations ($\gamma_{++}$ and $\gamma_{--}$) are sensitive to the presence of the CME; charge flow along the collision magnetic field vector or opposite to the direction of that vector causes $\gamma_{++}$ and $\gamma_{--}$ to be negative and $\gamma_{+-}$ to be positive.   However, $\gamma$ is also sensitive to background correlations unrelated to the CME, most significantly from multi-particle correlations within the event and two-particle correlations that vary in strength depending on azimuthal angle.  

The STAR collaboration measured $\gamma$ in heavy ion collisions and found a signal consistent with some expectations for the CME \cite{Abelev:2009ac}.   Effects other than the CME have been proposed as explanations for this signal \cite{Schlichting:2010qia,Bzdak:2010fd}, and subsequent work has included attempts to better define expectations from the CME within a heavy-ion event \cite{Yin:2015fca}, measurements of $\gamma$ for other collisions energies \cite{Adamczyk:2014mzf,Abelev:2012pa}, and measurements of observables sensitive to other effects of related origin that may be present in heavy-ion collisions (the "Chiral Magnetic Wave" \cite{Kharzeev:2010gd,Adamczyk:2015eqo} and "Chiral Vortical Effect" \cite{Son:2009tf}).   The current state of understanding on this topic was summarized by a recent task force appointed by the management of Brookhaven National Lab.  They concluded:  "A measurement of charge separation in heavy-ion collisions that can be unambiguously linked to the chiral magnetic effect would be of great interest to the wider physics community and would contribute significantly to the scientific impact and legacy of RHIC.  Many measurements have been carried out to study charge separation in heavy-ion collisions that are generally in agreement with expectations from the CME.  Background models however, can also account for much of the data. Based on our current understanding, backgrounds may account for all of the observed charge separation." \cite{Skokov:2016yrj}.

In this paper, we propose an additional experimental probe for LPV.  It is the measurement of event-by-event correlations between the charge separation along the magnetic field and the net helicity of $\Lambda$ and 
$\bar{\Lambda}$ particles produced in the event.  As we will argue, there are clear expectations for the sign of these correlations due to LPV effects.  With a simple model of the collision, we will estimate the size of the correlation signal that is expected in light of previous measurements.

\section{Proposed Measurement }

At the root of the Chiral Magnetic Effect is a net chirality of the event caused by the topological charge, $Q$, associated with that event.  These are related by $2Q = N^{f}_{L} - N^{f}_{R}$ where $N_L$ ($N_R$) denotes the number of quarks plus antiquarks with left-handed (right-handed) chirality and $f$ represents a particular flavor of quark.  Importantly, the same $Q$ applies to all "light" quark flavors, so in the limit that $u$, $d$, and $s$ quarks may all be treated as massless, we have
\begin{equation}
\label{eq:Q}
(N^{u}_{L} - N^{u}_{R}) =  (N^{d}_{L} - N^{d}_{R}) = (N^{s}_{L} - N^{s}_{R}) = 2Q
\end{equation}
The extent to which the strange quark may be treated as chiral is an open question \cite{Kharzeev:2010gr}, and in fact recent calculations in \cite{Mace:2016shq} indicate that the reduction of this effect among strange quarks due to their larger mass will be significant.  For the model calculations in Section \ref{sec:ModelCalculations} it is assumed that $u$, $d$, and $s$ quarks all follow Equation \ref{eq:Q}.

It is argued in \cite{Kharzeev:2007jp} that in the presence of the large magnetic field caused by the colliding ions, one effect of this net chirality is a net flow of electric charge along the direction of the generated magnetic field (this is the CME).  In events with $Q<0$, the CME effect should cause a flow of positive charge in the direction of the magnetic field, negative charge in the opposite direction.  In events for which $Q$ is greater than zero, the net charge flow due to the CME will reverse.  

The magnetic field direction in a heavy-ion event can be estimated from measurements of the 1st order reaction plane due to azimuthal asymmetry of particle production as a function of rapidity.  The 1st order reaction plane has generally not been used in previous studies of the Chiral Magnetic Effect (a notable exception is in \cite{Adamczyk:2013hsi}), because only the 2nd order reaction plane is needed to construct the observable $\gamma$.  The first order plane is necessary for the measurements proposed here. 

The net chirality given to strange and anti-strange quarks should have other observable effects besides the CME.  It should also result in a net helicity of produced $\Lambda$ particles.  In the simple quark model of hyperons, the  $\Lambda$ spin is completely determined by the spin of its strange quark \cite{Burkardt:1993zh}.  Other calculations give the fraction of $\Lambda$ spin carried by the strange quark as closer to 70\% \cite{Gockeler:2002uh} and analysis of experimental results indicate a value somewhere between those two \cite{Ellis:1995fc}.  The important point is that an effect of LPV within the event should be a net helicity of produced $\Lambda$ particles, with the sign and magnitude of the net helicity being governed by the same $Q$ which leads to charge separation.

To summarize: in events with a negative value of $Q$, the LPV effects should include a positive charge flow in the direction of the magnetic field {\it and} an excess of $\Lambda$s with right-handed helicity compared to the number with left-handed helicity.  In an event with a positive $Q$ value, the sign of both of these effects should be reversed; a net flow of negative charge in the magnetic field direction and an excess of left-handed $\Lambda$s.  We expect then an event-by-event correlation of the signs and magnitudes of these effects if LPV is present, with a clear expectation for the sign of the correlation.  A way to study this would be to look at the net helicity for all identified Lambdas in an event, $N(\Lambda_{RIGHT}) - N(\Lambda_{LEFT})$, versus the charge separation among pions in that event, $N(\pi^+_{up})  - N(\pi^-_{up})  - N(\pi^+_{down})  + N(\pi^-_{down})$, where "$up$" is defined as being in the direction of the magnetic field vector found via the 1st order reaction plane.  The expected LPV signal is a positive correlation between these two quantities.

Another important prediction is that LPV will tend to cause the same sign of net $\bar{\Lambda}$ helicity and net $\Lambda$ helicity in a given event, so that the event-by-event net helicities of both $\Lambda$ and $\bar{\Lambda}$ should give the same sign of correlation with the charge separation.  The purpose of this paper is to point out that such correlations should exist in the presence of LPV and that measurement of them would give another experimental handle on chiral effects in heavy-ion collisions.

\section{Estimation of Correlation Strength in Heavy-Ion events}\label{sec:ModelCalculations}

With the simple simulation model described below, we will make a rough estimate of how large a correlation signal may be present in heavy-ion events, and how many events may be needed to see such a correlation above statistical fluctuations.  We assume in this model that we are looking at minimum-bias Au-Au collisions with energy $\sqrt{s_{NN}}=$200GeV/c.  We treat each centrality bin separately (using bins of 10-20\%, 20-40\%, and 40-60\% and assuming there is no appreciable signal in the other centrality bins), but when discussing numbers within the model we will for concreteness refer to the case of 20-40\% centrality.  Motivated by the capabilities of the STAR detector, we assume an acceptance in rapidity of $\Delta y = 1$.

LPV in an event should have the effects of charge separation along the magnetic field and a net helicity of both $\Lambda$s and $\bar{\Lambda}$s.  The strength of all these effects in a given event is governed by the value of $Q$ in the event.  To estimate a correlation strength we must make some assumption about the event-by-event distribution of values of $Q$.  We assume a Gaussian distribution (though these results are rather insensitive to the shape of the distribution) with a mean of zero and a width of the distribution $\sigma_Q = 1.4 $ that reproduces the charge separation signal from \cite{Abelev:2009ac}.  This may seem rather liberal, essentially assuming that all the observed charge separation signal is due to the CME, but we note that it is also possible that some of the initial CME signal is dissipated in the evolution of the collision and therefore the charge separation may not reflect the full value of $Q$ that could affect the hyperon helicity.  

In our model, to simulate a charge separation signal for an event, we start by choosing a value $Q_{evt}$ for the topological charge of that particular event and then generating $\pi^+$ and $\pi^-$ particles with multiplicities as measured in \cite{Abelev:2008ab} and random in azimuthal angle (with the magnetic field direction always chosen to be at $\pi/2$ radians).  For each increment of $Q$ in the particular event, we then reverse the momentum of one charged pion; so for an event with $Q_{evt}=-2$ a possibility would be to reverse the momentum one $\pi^+$ from heading "down" with respect to the magnetic field ($-\pi/4 < \phi < -3\pi/4$) to heading "up" (by adding $\pi$ to $\phi$) and to reverse the momentum of one $\pi^-$ from heading "up" to heading "down".  We verify that this crude process with the parameters we have chosen closely reproduces the charge separation signal reported in \cite{Abelev:2009ac} for $\gamma_{++}$ and $\gamma_{+-}$.

To simulate the effect of LPV on $\Lambda$ helicities (and $\bar{\Lambda}$ helicities; in the following paragraphs, we use "$\Lambda$"  and "$s$ quarks" to mean both particles and antiparticles) we assume that the number of $s$ quarks that undergo a change in helicity is equal to $Q_{evt}$ as implied by Equation~\ref{eq:Q} under the assumption that the strange quark may be considered a light quark flavor.  We furthermore assume that each $\Lambda$ helicity is completely determined by the helicity of its strange quark.  For the probability that any particular strange quarks becomes a constituent of a "primary" $\Lambda$, we take the number of produced $\Lambda$s divided by the total number of strange quarks present in final-state particles.  These numbers we take from previous measurements \cite{Adams:2006ke,Agakishiev:2011ar,Abelev:2008ab} and find the the probability is $p_{\Lambda}=0.146$.  As noted below, "primary" $\Lambda$s in our model include feed-down from $\Xi$ baryons but not from $\Sigma^{0}$.

In our model, we generate in each event a number of $\Lambda$s consistent with \cite{Agakishiev:2011ar}, each with random helicity, and then for each integer increment of $Q_{evt}$, there is a probability $p_{\Lambda}$ that one of the $\Lambda$ helicities is flipped; from $R$ to $L$ if $Q_{evt}$ is positive, from $L$ to $R$ if $Q_{evt}$ is negative.  For simplicity, we make the assumption that secondary $\Lambda$s that are daughters of $\Sigma^{0}$ hyperons (25\% of total $\Lambda$s) do not carry any effect from strange quark polarization, but that $\Lambda$s that are daughters of $\Xi$ baryons carry the same effect as primordial $\Lambda$s.   This latter assumption is motivated by the constituent quark model in which the  $\Xi$ spin is highly correlated with the spin of its strange quark along with the fact that the polarization of the $\Xi$ is largely transferred to its daughter $\Lambda$ \cite{Becattini:2016gvu}.

Finally, we fold in experimental assumptions driven by the performance of the STAR detector - an efficiency of $\epsilon_{\Lambda}$=5\% for detecting each $\Lambda$ (also included is the probability that the helicity will be correctly determined by the daughters' direction in its parity-violating decay) and a first order event plane resolution of 0.4 that is implemented by rotating all azimuthal angles in a given event by the same randomly chosen amount.  For each event we then have two quantities: one is the net helicity for all identified Lambdas, $H = N(\Lambda_{RIGHT}) - N(\Lambda_{LEFT})$, the other is the charge separation $C = N(\pi^+_{up})  - N(\pi^-_{up})  - N(\pi^+_{down})  + N(\pi^-_{down}) $ where "$up$" is defined with respect to the 1st order reaction plane.  The expected LPV signal is a positive correlation between these two quantities.

We quantify this correlation by calculating the sample Pearson correlation coefficient, $r_{H,C} = (\sum_i H_i C_i - n \bar{H} \bar{C})/s_H s_C$ with the sum and averages calculated over the simulated sample of $n$ events for which the quantities have sample standard deviations $s_H$ and $s_C$.  Under the assumption that arctanh(r) follows a normal distribution with variance $1/n$ , we determine the number of events needed so that if there were no correlation between $H$ and $C$, there would be a $<5\%$ chance of a positive correlation value as large as the one seen in the simulated data. 

\begin{figure} 
\includegraphics[width=\textwidth]{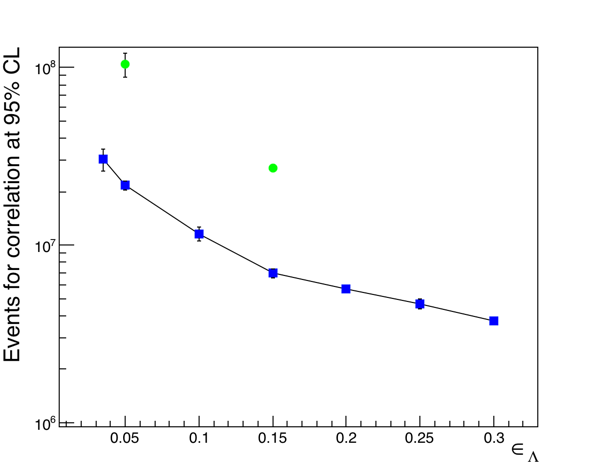}
  \caption{Toy model calculation of the number of minimum-bias events required to see a correlation between event-by-event net $\Lambda$ helicity and charge separation with respect to the 1st order reaction plane.  Blue squares represent the assumption that all charged particle correlation signal is due to the CME and assumes both $\Lambda$ and $\bar{\Lambda}$ are used, with the calculation given as a function of experimental efficiency for finding $\Lambda$s varying from 3.5\% to 30\%.  Green circles represent the number of events needed assuming only 1/2 of the measured charged particle correlation signal comes from CME.  Statistical uncertainties from the simulations are shown on all points. \label{fig:Figure1}}
\end{figure}

Our simulation gives a result of 22 million for the number of 200 GeV minimum bias events necessary to observe a positive correlation between charge separation and net $\Lambda$+$\bar{\Lambda}$ helicity at a $95\% $ confidence level as defined above.  This is shown as the blue square corresponding to $\epsilon_{\Lambda}$=5\% in Figure \ref{fig:Figure1}.  To illustrate the dependence of the size of the measured correlation on experimental efficiency, we have also done this calculation for $\epsilon_{\Lambda}$ values in the range from 3.5\% to 30\% and these results are displayed as the blue squares in Figure \ref{fig:Figure1}.  Statistical error bars are included on all points.  For reference, STAR has recorded over 1 billion minimum-bias Au+Au events at $\sqrt{s_{NN}}=$200GeV/c.

The green circles in Figure \ref{fig:Figure1} represent a more conservative assumption regarding the observed charged particle correlation signal.  For these points, rather that assuming a distribution of $Q$ values that would be consistent with the CME causing the entire signal measured in \cite{Abelev:2009ac}, we assume that the CME is responsible for one-half of the charge-separation signal, with the rest coming from small-angle clusters of two opposite-signed particles.   Under this assumption, our simulation model indicates about 100 million events would be needed to see a correlation at 95\% CL assuming $\epsilon_{\Lambda}$=5\%. 

A still better test, though more statistically challenging, would be to look for correlations of $\bar{\Lambda}$ helicity with charge particle separation separately from the correlation of $\Lambda$ helicity with charge particle separation.  From LPV effects, both of these should show the same sign of correlation.  From our model, the number of events to see the two correlations separately can be estimated by simply varying the value of $\epsilon_{\Lambda}$.  For example, to see separately the $\bar{\Lambda}$ {\it or} $\Lambda$ correlations in a detector with $\epsilon_{\Lambda}$=10\% requires approximately the same number of events as to see the combined $\bar{\Lambda}$ {\it and} $\Lambda$ correlations in a detector with $\epsilon_{\Lambda}$=5\%.

This model deals only with the statistics necessary to see a correlation signal and does not deal with the systematics involved in making the measurement.  One potential issue is that due to decay kinematics and detector acceptance, there may be very different efficiencies for $\Lambda$s of different helicities.  For example the STAR detector has much larger efficiency for $\Lambda$s with right-handed helicity than with left-handed helicity (with the opposite true for $\bar{\Lambda}$s ) \cite{STAR:2017ckg}.  Within our simulation, differing efficiencies for left- and right-handed helicities did not significantly degrade the correlation signal.  

\section{Summary}

In this paper, we have advocated for the idea of measuring event-by-event correlation between $\Lambda$ and $\bar{\Lambda}$ helicity and charge separation along the magnetic field direction (as indicated by the {\it first} order reaction plane) as a probe of Local Parity Violation and the Chiral Magnetic Effect in heavy-ion collisions.  The sign of the correlation for both $\Lambda$ and $\bar{\Lambda}$ due to LPV has a clear expectation.  Such a signal should not be susceptible to the flow-related backgrounds that muddy the interpretation of the charged-particle correlation alone.  

Although several assumptions in our toy model are likely at least slightly optimistic and one (treating the strange quark as chiral) has a large uncertainty, results from it indicate that if the charge separation signal measured in \cite{Abelev:2009ac} is largely the result of CME effects, there is a reasonable chance of observing the combined $\Lambda + \bar{\Lambda}$ signal in data already taken at RHIC.  If data from the isobar running at RHIC in 2018 leads to successfully establishing that measured charged correlations are due to CME \cite{Deng:2016knn}, it would then be interesting still to search for such correlations to see the LPV effect on particles' helicity and perhaps as a useful probe of the dissipative effect of the non-zero strange quark mass.



%


\bibliography{PAPER}

\end{document}